\documentstyle[prl,twocolumn,aps,epsf]{revtex}
\begin{document}
 
\twocolumn[\hsize\textwidth\columnwidth\hsize\csname
@twocolumnfalse\endcsname
\title{Fullerene-based molecular nanobridges: A first-principles study}
\author{ J. J. Palacios, A. J. P\'erez-Jim\'enez, and E. Louis}
\address{Departamento de F\'{\i}sica Aplicada y Unidad Asociada UA-CSIC,
Universidad de Alicante, San Vicente del Raspeig, Alicante 03080, Spain}
\author{J. A. Verg\'es}
\address{Instituto de Ciencia de Materiales de Madrid (CSIC),
Cantoblanco, Madrid 28049, Spain.}
\date{\today}
\maketitle
  
\widetext          
\begin{abstract} 
Building upon traditional quantum chemistry calculations, we have
implemented an {\em ab-initio} method to study the electrical
transport in nanocontacts. We illustrate our technique calculating the
conductance of C$_{60}$ molecules connected in various ways to
Al electrodes characterized at the atomic level.  Central to a correct
estimate of the electrical current is 
a precise knowledge of the local charge transfer between molecule and metal
which, in turn, guarantees the correct positioning of the Fermi level with
respect to the molecular orbitals. 
Contrary to our expectations, ballistic transport seems 
to occur in this system.
\end{abstract} 

\vskip2pc]
 
\narrowtext

At the forefront of the carbon-based molecules that are expected to
play a key role in nanoelectronic devices one can situate carbon
nanotubes and fullerenes.  From Coulomb blockade to Kondo-like
phenomena through ballistic transport, multiple transport regimes have
been observed in nanotubes (even for the same chirality of the
nanotube) depending on the way the contact  with the electrodes is
made\cite{Dekker}.  Very little is known, however, about the true role
played by this contact as well as of the effect of the substrate on
which the carbon nanotube has been layed.  This is somewhat connected
with the difficulties experienced in the interpretation of electronic
transport experiments through C$_{60}$ molecules.
Experiments using a scanning tunneling microscope
(STM)\cite{stm} have shown the possibility
of ``looking" at individual C$_{60}$ molecules when they are adsorbed
on a conducting substrate. There is still, however, a lot of
controversy in the interpretation of the images obtained by different
groups\cite{Julio_comment}.  Alternative to STM set-ups,  nanoscopic
break junctions also revealed themselves as powerful tools to study
transport through individual molecules\cite{Reed,Kergueris}.
Recently, for instance, in between the two electrodes of a gold break
junction, evaporated  C$_{60}$ molecules
unexpectedly created ``mechanical'' bridges for electrical
transport between electrodes\cite{Park}.

Archetypal STM and break-junction set-ups above mentioned
can be cataloged under the term molecular nanobridge, i.e.,
metallic electrode + molecule + metallic electrode. The conductance,
$G$, is essentially
determined by the electronic structure of the isolated molecule since
the number of available channels for conduction deep in the electrodes
is always much larger than those provided by the molecule.  However,
as already mentioned, this conductance is strongly influenced by the
chemistry of the electrode-molecule contacts.  Behind this contact
lies a key problem: Charge transfers between electrode and molecule
when they come in close proximity.  The necessity of aligning the
Fermi level of the metallic electrodes with the ``Fermi level'' of
the molecule forces a certain amount of charge to be transferred one
way or the other. The way this alignment is achieved and how much
charge is transferred constitutes a difficult problem which turns out
to be essential to understand the electrical transport in molecular
nanobridges and, more generally, in nanocontacts or nanoconstrictions.
Curiously enough, the prevailing theoretical descriptions in the great
body of work done on various types of metal-molecule-metal
systems\cite{Joachim_00} as well as on atomic chains and
constrictions\cite{Cuevas} are still based on parametrized
tight-binding or semi-empirical models.  The theoretical framework to
calculate the current is well established\cite{Datta}, but,   within
these models, Fermi level alignment has to be either entirely ignored
or imposed with some additional criteria.  While local charge
neutrality is a sensible criterion for atomic chains and constrictions
made of the same material as the electrodes\cite{Cuevas}, there is not
the like for metal-molecule-metal systems. This is, in  part, the
motivation for developing {\em ab-initio} methods which can provide
qualitative and quantitative answers in all possible scenarios.  Some
groups have already blazed the trail in this
direction\cite{Lang,Guo,Mujica,Guo_preprint}, although in their
methods the problem of the atomic structure of the electrodes is
usually put aside.  This is not satisfactory when one is trying to
describe STM experiments where the detailed atomic structure of the
tip determines, to a large extent, whether or not the STM images can
resolve the topography or molecular structure of the adsorbate. 

What we present here is an {\em ab-initio} alternative to parametrized
tight-binding or semi-empirical models in order to address electrical
transport through generic nanocontacts. Our method automatically
accounts for Fermi level alignment and charge transfer whenever
the latter is expected to occur. This is the case in the system 
we study here: Al$_{\infty}$-C$_{60}$-Al$_{\infty}$.
Isolated C$_{60}$ molecules present a large gap ($\approx$ 2 eV.) and
behave like intrinsic semiconductors. At low temperatures,
metal-semiconductor-metal systems {\em do not} conduct when the
semiconductor is intrinsic
regardless of the sign and the amount of charge transferred at the
interfaces.  In the metal-fullerene-metal system, however, a tunnel
current can always be expected due to the nanoscopic size of the
fullerene\cite{Joachim_98}.  What is more surprising is the fact that 
C$_{60}$ molecules can conduct ballistically 
($G>e^2/h$)\cite{Guo_preprint}.  Some of
the lowest unoccupied molecular orbitals (LUMO's) of the isolated
fullerene still retain its extended character when contact is made to
the electrodes and they become partially or fully occupied. 
In solid-state language:
The Fermi level lies above the bottom of the conduction band all
across the (nanoscopic) semiconductor. Notice that bigger molecules
exhibiting a gap like some families of undoped nanotubes are not expected to
conduct at zero bias\cite{Dekker}. Whether or not this
occurs depends on the size of the molecule and
the details of the charge transfer process which, in
turn, relies heavily on the relaxation of the atomic structure of the
fullerene when in contact with the metal.

\begin{figure}
\vspace{-0.3cm}
\centerline {\epsfxsize=7cm \epsfysize=4.3cm \epsfbox{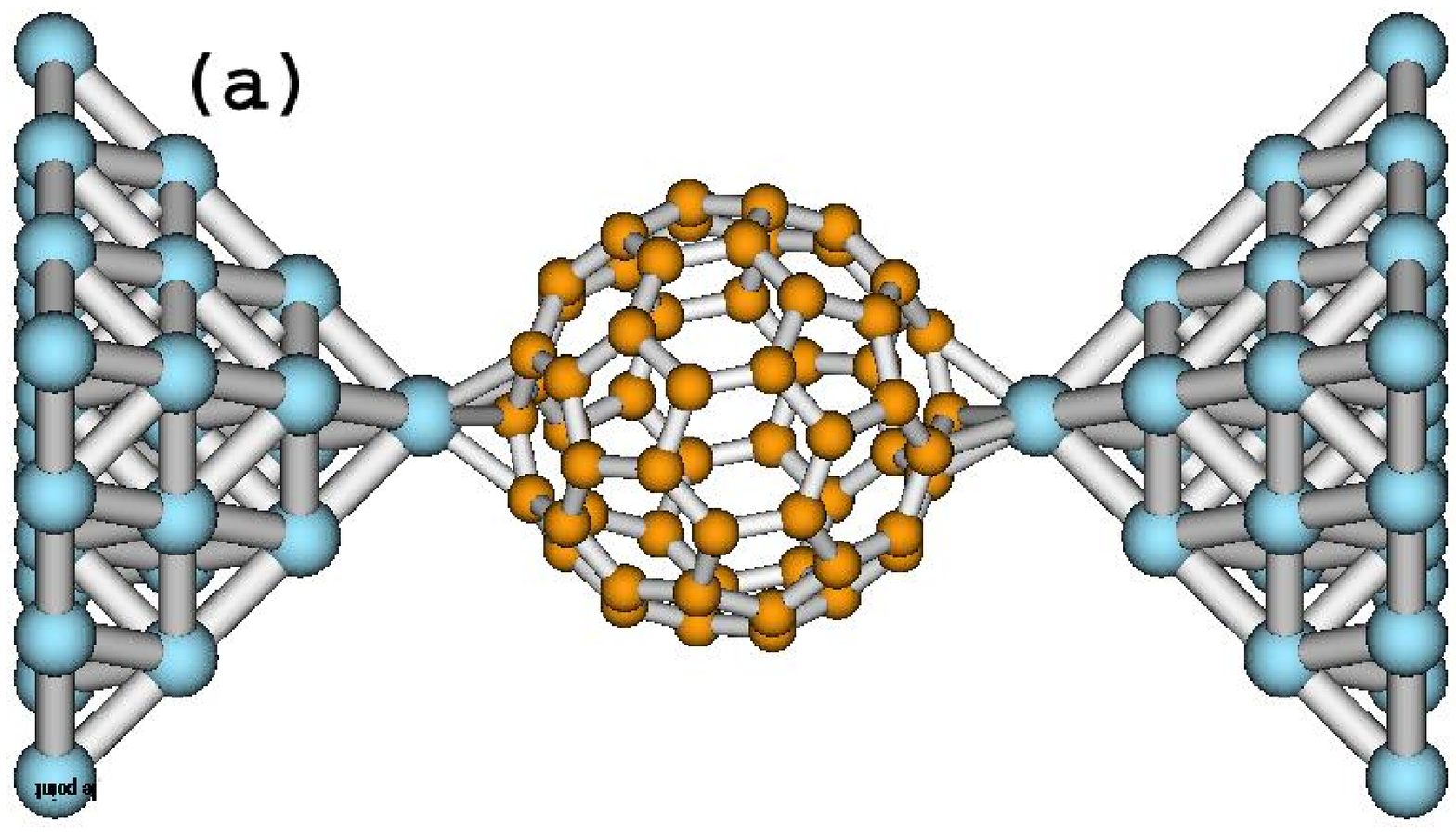}}
\vspace{-0.4cm}
\centerline {\epsfxsize=7cm \epsfysize=3.5cm \epsfbox{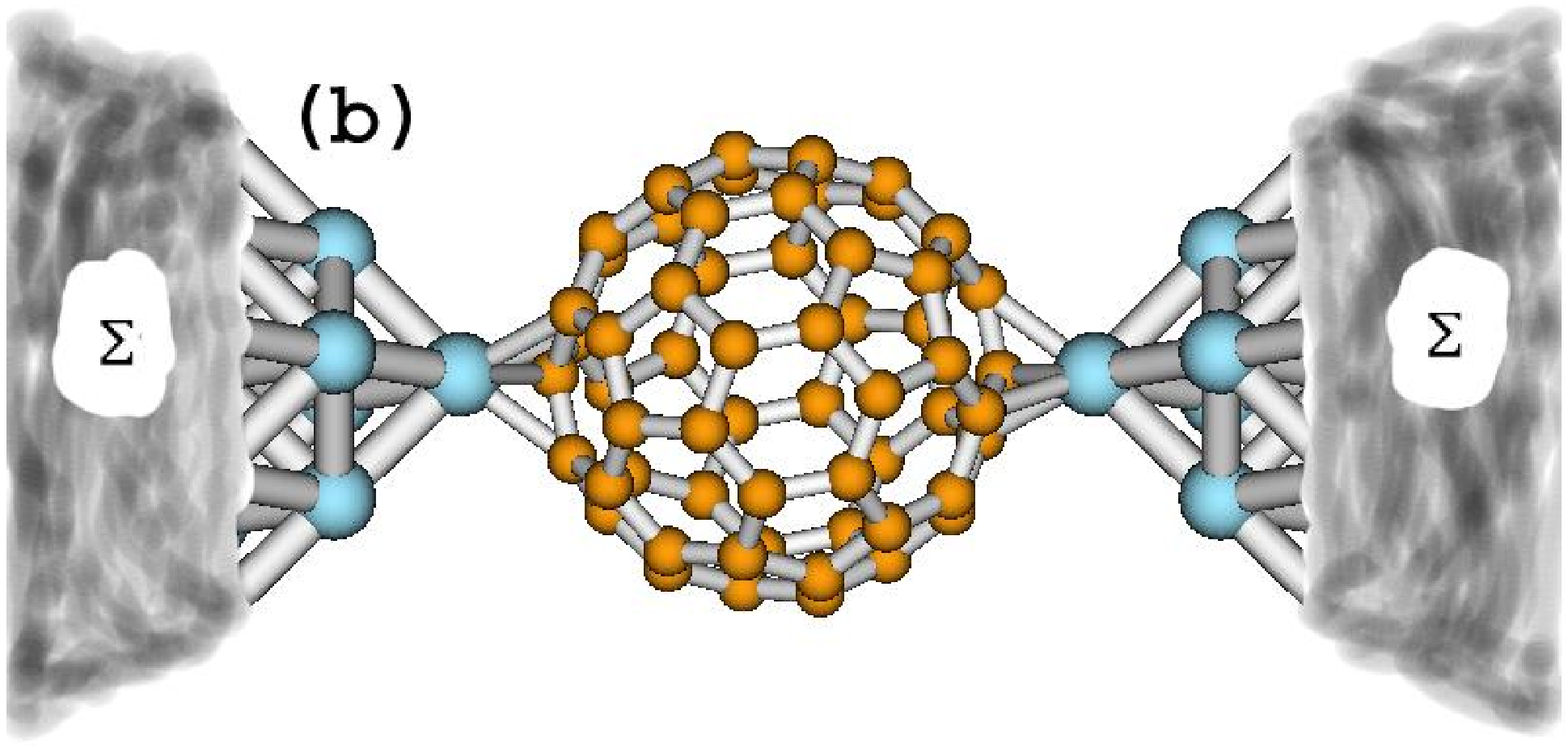}}
\caption{Schematic view of the two-step process for calculating the
Green's
function used in the calculation of the conductance:
(a) A density functional calculation of a
C$_{60}$ molecule contacted by two
electrodes (here in the form of pyramids) is performed.
(b) Atoms that are irrelevant for electrical 
transport are removed and substituted by an 
effective selfenergy ``attached'' to the contact left and right atoms 
remaining in the electrodes (see text).}
\label{pyr}
\end{figure}     
Standard {\em ab-initio} or 
quantum chemistry calculations\cite{Gaussian,Ordejon}
are only possible for finite or periodic systems, whereas the 
transport problem one typically wants to
address requires infinitely large leads with no symmetry at all. 
Drawing on Refs. \onlinecite{Mujica,Guo_preprint},
the main idea here consists of performing a density functional (DF) 
calculation
of the molecule {\em including part of the leads with the desired
geometry} [see Fig. \ref{pyr}(a)].  This can be efficiently done
using the Gaussian-98 code\cite{Gaussian}.
For a sufficiently large number of atoms describing the metallic electrodes
the energy of the highest occupied molecular orbital (HOMO) 
of the whole structure sets the Fermi level. In addition,
the correct charge transfer (to or from the molecule) is guaranteed.
Next, we perform a L\"owdin orthogonalization of the original
non-orthogonal gaussian basis set into an orthogonal one which,
having basis elements more extended than the original one, still preserves
the symmetry of the orbitals and the atomic character. 
The hamiltonian of the electrode+molecule+electrode system, $\hat{H}$, 
as it stands, is finite and, according to the usual theoretical transport
schemes\cite{Datta}, its Green's functions are unsuitable for 
any current determination since they simply have poles.  The following
(and crucial) step in our procedure is to transform this finite system into an
effectively infinite one. In order to do this we first remove from
the hamiltonian all but the $N$ atoms forming the relevant atomic
structure of each electrode close to the molecule [see Fig.
\ref{pyr}(b)].
The retarded Green's function associated with this
reduced hamiltonian $\hat{H}_r$
is now transformed into the retarded Green's function of an infinite system:
\begin{equation}
G^r(\epsilon)= (\epsilon\hat I  -  \hat H_r +i\delta)^{-1} \rightarrow 
[\epsilon\hat I  -  \hat H_r - \hat\Sigma(\epsilon) ]^{-1}.
\end{equation}
In this expression $\hat\Sigma=\hat\Sigma_R + \hat\Sigma_L$ where 
$\hat\Sigma_R$($\hat\Sigma_L$) denotes a self-energy matrix that  
accounts for the right(left) infinite electrode, part of which has
been included in the DF calculation [see Fig. \ref{pyr}(a)]. 
The added self-energy can only be explicitly calculated in 
ideal situations, which, in principle, might seem to limit the desired 
applicability of the above step. However, as pointed out by
Landauer in his scattering description of electrical
transport\cite{Datta}, 
the bulk details of a metallic electrode are not necessarily
relevant for the electrical transport properties of the region with the 
smallest number of channels, i.e., the contacted molecule.
With this crucial observation in mind we choose to describe the bulk electrode
with a Bethe lattice tight-binding model\cite{Martin} 
with the appropriate coordination
and appropriate parameters. More specifically: 
We require  our Bethe lattice model  to reproduce
the electrode bulk density of states and to have the same 
Fermi energy as that of the
system on which the DF calculation was initially performed\cite{note1}.
The advantage of choosing a Bethe lattice resides in that $\hat\Sigma$ 
can be  easily
calculated through a well-known iteration procedure\cite{Martin} 
which we do not detail here. It suffices to say that, for each atom 
in the reduced electrode that has been stripped from, at least, one 
nearest-neighbor atom, a contribution from an infinite Bethe lattice
calculated in the direction of the removed atom is added to
$\hat\Sigma$.  Assuming
that the initial system was large enough and that the reduced system still 
contains the relevant atomic details of the electrode 
close to the contact with the molecule, to replace a  piece of the
electrode by our effective self-energy
should introduce no spurious effects, {\em and it makes the system
effectively infinite}. The conductance can now be simply 
calculated through the expression\cite{Datta}
\begin{equation}
G=\frac{2e^2}{h}{\rm Tr}[\Gamma_L G^r \Gamma_R G^a],
\end{equation}
where Tr denotes the trace over all the orbitals 
of the reduced hamiltonian and the matrices 
$\Gamma_R$ and $\Gamma_L$  are given by $i(\Sigma^r_R-\Sigma^a_R)$ and 
$i(\Sigma^r_L-\Sigma^a_L)$, respectively. 
As in the self-energy matrices, the matrix 
elements of $\Gamma_R$ and $\Gamma_L$ are only different from zero for
the atoms in the reduced electrodes that have been stripped from at
least one near-neighbor atom.

\begin{figure}
\vspace{-0.5cm}
\centerline {\epsfxsize=8.5cm \epsfysize=6cm
\epsfbox{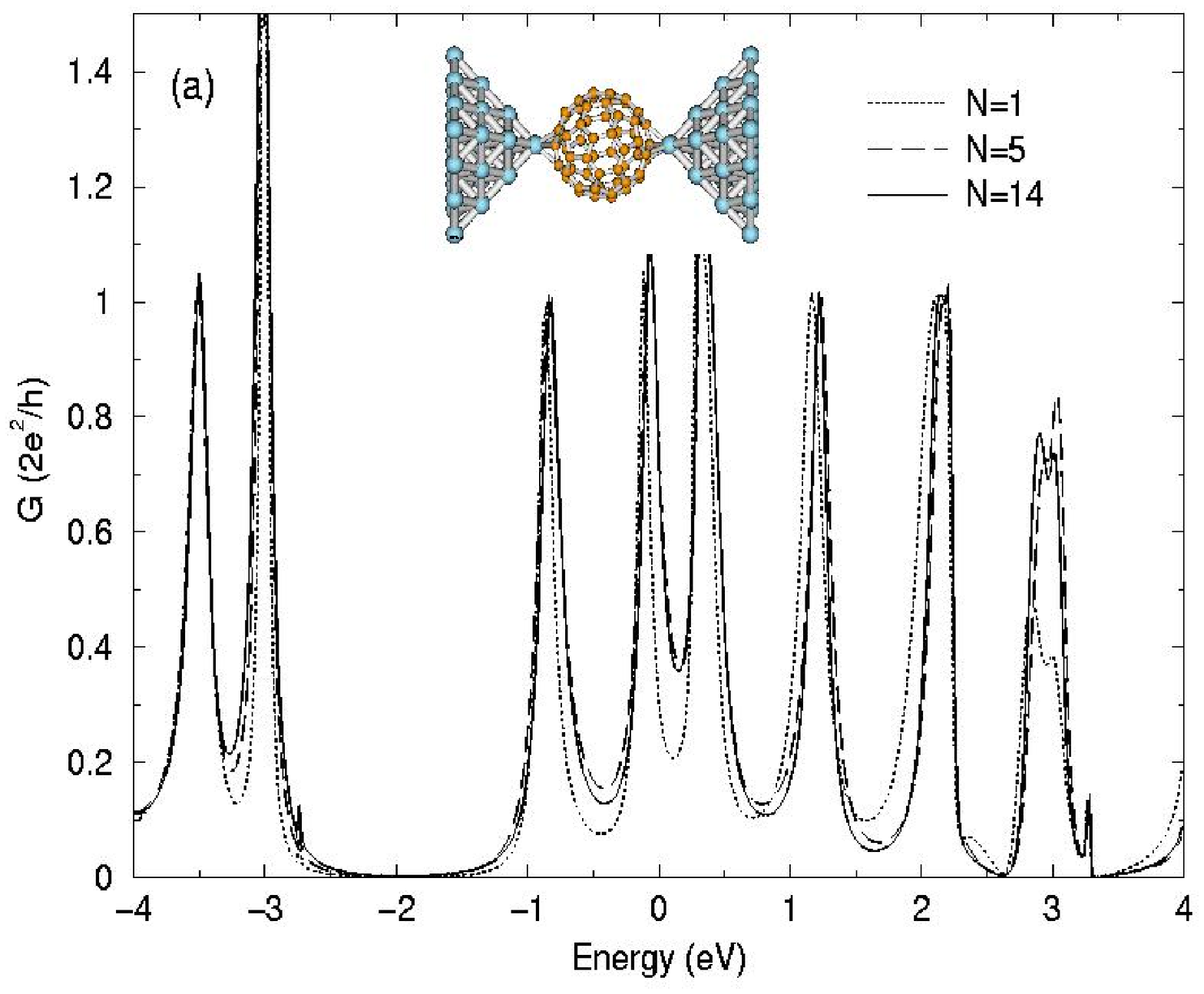}}
\vspace{-0.5cm}
\centerline {\epsfxsize=8.5cm \epsfysize=6cm
\epsfbox{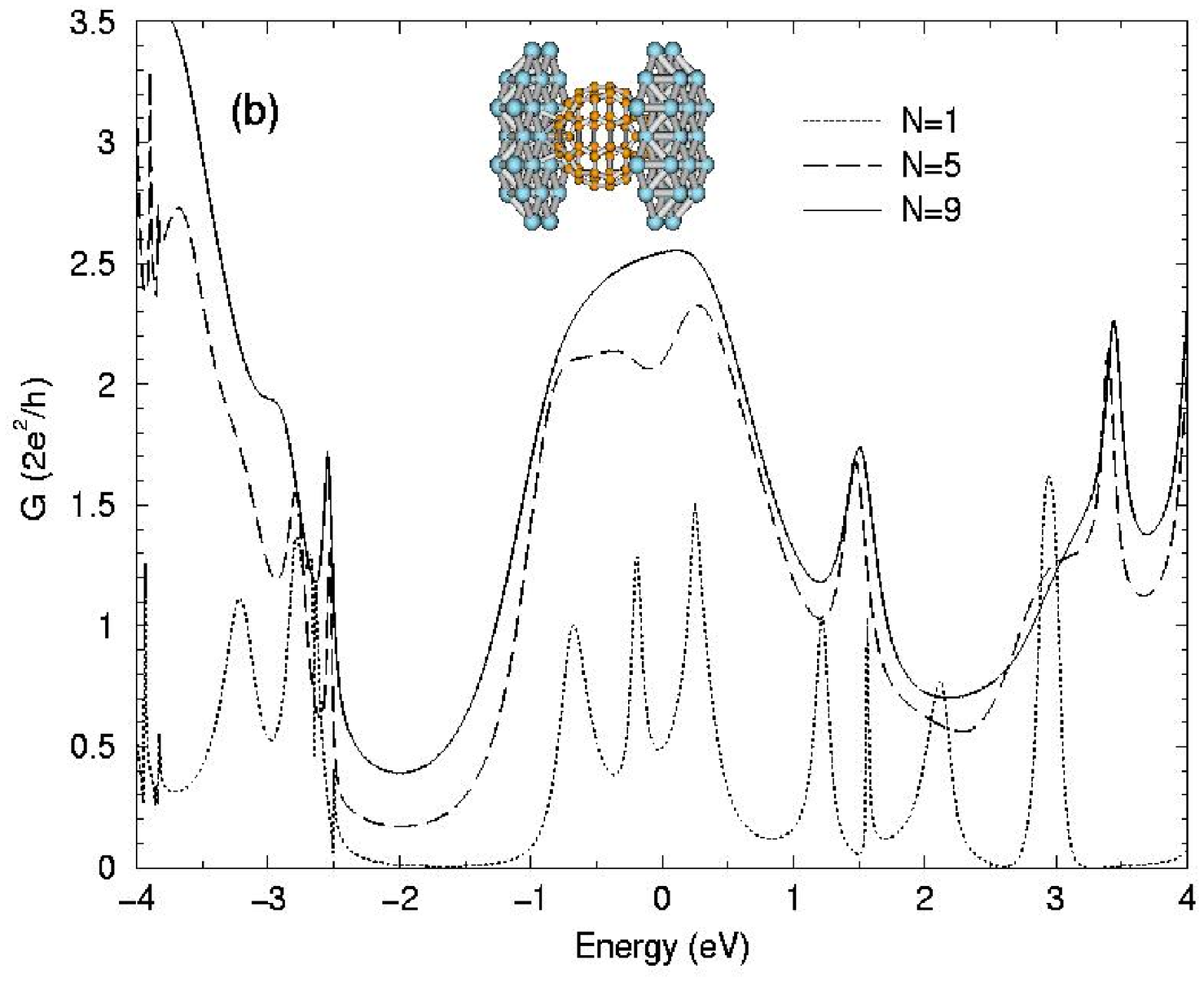}}
\caption{(a) Conductance around the Fermi energy (here set to zero) of a
C$_{60}$ molecule contacted by two opposite
Al electrodes in the form of pyramids composed of 30 atoms
each (see inset).  We present three different cases:
$N=1$ (dotted line), $N=5$ (dashed line), and $N=14$ contact atoms (solid
line), corresponding to an increasingly detailed electrode structure in the 
reduced system (see text). (b) Same as in (a), but with the fullerene
in between Al(001) parallel surfaces, each one composed  by two atomic
planes with a total of 33 atoms (see inset). We present results for 
$N=1$ (dotted line), $N=5$ (dashed line), and $N=9$ 
contact atoms (solid line), corresponding to an increasing
surface plane.}
\label{G}
\end{figure}  
We apply here the technique just described to the
system Al$_{\infty}$-C$_{60}$-Al$_{\infty}$ with different
geometries. Although they are not commonly used
in experiments, we have considered Al electrodes for 
two independent reasons: (i) The number of electrons per atom 
is small enough to avoid
pseudo-potentials in the DF calculations and, (ii) in contrast to noble
metals, Al presents a strong, essentially covalent chemical bond with carbon
molecules which, in principle, makes it ideal for the future fabrication of
stable nanostructures.  For the DF
calculation we have used the Becke's three-parameter hybrid functional
using the Lee, Yang and Parr correlation functional with a minimal basis
set\cite{Becke}.  The orientation of 
the fullerene and the separation and atomic structure of the electrodes
considered in  Fig. \ref{G} (see insets) are reasonable, 
but arbitrary\cite{note2}.  In Fig. \ref{G}(a) we plot the
conductance around the Fermi energy (here set to zero) of
the fullerene contacted by two opposite electrodes in the form of pyramids 
[see inset or Fig.\ref{pyr}(a)]. These are composed 
of 30 atoms each and their apex atoms are 
separated by 10.1 $\AA$ . 
We have  orientated the fullerene so
that two opposite C atoms are as close as possible to the 
apex atoms of the pyramids (i.e., 1.4 $\AA$ between the C atom of the fullerene 
and the Al apex atom). In order to illustrate
the reliability of the method, we present the results for three 
different approximation degrees:
$N=1$ (dotted line), $N=5$ (dashed line), and $N=14$ (solid
line), corresponding to an increasingly detailed contact structure.
Notice how, as expected, all the curves are very similar
since the relevant atomic structure close to the molecule is 
simply represented by the atom at the pyramid apex. Logically,
the results  using 5 atoms (two atomic planes)
are almost identical to those using 14 atoms (three atomic planes). 
The gap of the isolated fullerene reflects itself in a vanishing $G$
below the Fermi energy and the LUMO's of the isolated
molecule are now partially occupied. Note that they retain their extended
character since the conductance around the Fermi level goes up to 
$2e^2/h$ at certain energies. Nonetheless, the three-fold degeneracy of the
$f_{1u}$ and $f_{1g}$ LUMO's has been entirely removed by the
interaction with the electrodes
(this interaction is also responsible for the broadening of the
peaks). In this situation Coulomb blockade physics must dominate and 
spin-splitting will occur in the fullerene\cite{note3}.

In Fig. \ref{G}(b) we present the conductance of the fullerene oriented as in
the previous case, but where the two pyramids have been replaced by 
Al(001) parallel surfaces (see inset). The different
types of lines correspond to  an increasing definition of the 
surface plane (1, 5, and 9 surface atoms). 
Now one atom is not enough to define the contact since there is
an important coupling between several Al surface atoms and the fullerene. Not
even 5 atoms suffice
to define the metallic contact region quite precisely. Now the value of the
conductance rises up to $\approx 2.5e^2/h$ right at the Fermi energy. This was
somehow expected since, in this situation, the hybridization with the
surface is stronger and there are more channels for
the electron to travel between electrodes while the LUMO's appear to
be also extended. The gap still reflects itself in $G$, but has been
partially closed by the strong hybridization. The excess charge in
the fullerene is $\approx 2.6$ which agrees fairly well with a related and
recent calculation\cite{Guo_preprint}.
\begin{figure}
\centerline {\epsfxsize=8.5cm \epsfysize=6cm
\epsfbox{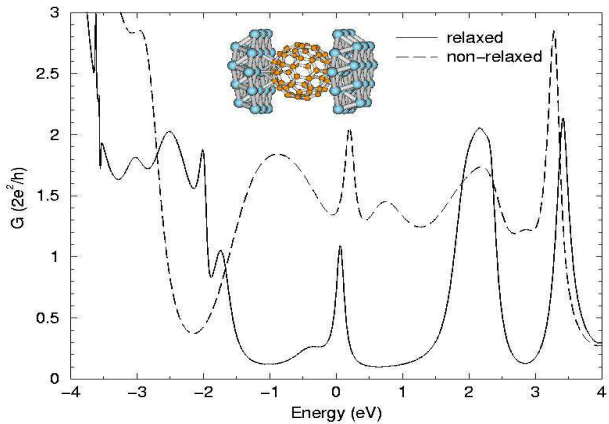}}
\caption{Conductance around the Fermi energy (here set to zero) of a
C$_{60}$ molecule contacted by two parallel Al(111) surfaces
separated by 10.1$\AA$ and composed
of 31 atoms each (see inset).  The dashed line
is the result for the non-relaxed fullerene and the solid one is the
result when the fullerene has been allowed to relax.}
\label{111}
\end{figure}  
Figure \ref{111} shows the conductance when the fullerene is placed in
between two parallel Al(111) surfaces (31 atoms each, 7 of which have
been used in the calculation of $G$) with parallel Al surface and fullerene
hexagons. This orientation has been chosen on the
basis of several experimental observations\cite{Fasel}.  In one case
(dashed line) the fullerene has not been allowed to relax. In the
other (solid line) an {\em ab-initio} relaxation has been performed.
The magnitude of $G$ in the former case is comparable to that shown in
Fig.  \ref{G}(b). Again, a large number of surface and fullerene atoms
intervene. While the details change, the gap is still visible around
-2 eV and the excess charge is $\approx 2.1$.  Surprisingly enough, when
the fullerene is allowed to relax (stretch in this case), the
conductance changes drastically. Only one LUMO around the Fermi level
preserves its extended character, the rest becoming localized at the
interfaces.  Both a Mulliken population analysis and an integration of
the local density of states show that the excess transferred charge
($\approx 3.3$) concentrates now more at the interface hexagons where
bonding takes place.  (The distance between surfaces was
chosen to be slightly larger than the diameter of the fullerene so
the structural changes of the fullerene do not play any role).
The small value of the
conductance below the Fermi energy cannot be attributed to the gap of
the fullerene (which has entirely disappeared from the density of
states), but to the localization process which does not reflect in
the global density of states of the fullerene. 
This last result bridges the gap with  the intuitive
picture for transport in a metal-semiconductor-metal system and
shows that the chemical details of the metal-molecule contact
are determinant in the transport properties of molecular bridges.

In summary, we have developed a methodology to
calculate {\em ab-initio} transport properties of nanoconstrictions. 
We have chosen to study a system of recent interest:
A fullerene contacted by Al electrodes. Our
main goal was to describe correctly the contact charge transfer
so that realistic predictions on the
electrical conduction properties of these molecules can be made.

Part of this work was supported by the 
Spanish CICYT under Grants Nos. 1FD97-1358
and by the Generalitat Valenciana under Grant No. GV00-151-01.
Discussions with E. San Fabi\'an, J. C. Sancho, L. Pastor-Abia, and J.
M. P\'erez-Jord\'a are acknowledged.


\begin{references}
\bibitem{Dekker}For a nice review on nanotubes see
C. Dekker, Phys. Today {\bf 52}, No. 5, 22 (1999).
\bibitem{stm}C. Joachim {\em et al.}, Phys. Rev. Lett. {\bf 74}, 
2102 (1995); D. Porath {\em et al.}, Phys. Rev. B {\bf 56}, 9829 (1997); J. G.
Hou {\em et al.},  Phys. Rev. Lett. {\bf 83}, 3001 (1999); J. I.
Pascual {\em et al.}, Chem. Phys. Lett. {\bf 321}, 78 (2000).
\bibitem{Julio_comment} J. I. Pascual {\em et al.},  Phys. Rev. Lett.
{\bf 85}, 2653 (2000); J. G. Hou {\em et al.},  Phys. Rev. Lett.
{\bf 85}, 2654 (2000).
\bibitem{Reed}M. A. Reed {\em et al.}, Science {\bf 278}, 252 (1997).
\bibitem{Kergueris}C. Kergueris  {\em et al.},  Phys. Rev. B {\bf 59},
12505 (1999).
\bibitem{Park} H. Park {\em et al.}, Nature {\bf 407}, 57 (2000).
\bibitem{Joachim_00}For references see
C. Joachim, J. K. Gimzewski, and A. Aviram, Nature {\bf 408}, 541 (2000).
\bibitem{Cuevas} J. C. Cuevas  {\em et al.},  Phys. Rev. Lett. {\bf
80}, 1066 (1998).
\bibitem{Datta} S. Datta, {\em Electronic transport in mesoscopic
systems}, ed. by H. Ahmed, M. Pepper, and A. Broers (Cambridge
University Press, Cambridge, 1995).                   
\bibitem{Lang} N. D. Lang and Ph. Avouris, Phys. Rev. Lett.
{\bf 81}, 3515 (1998).
\bibitem{Guo} G. Taraschi  {\em et al.},  Phys. Rev.
B {\bf 58}, 13138 (1998).
\bibitem{Mujica}S. N. Yarilaki {\em et al.}, J. Chem. Phys. {\bf
111}, 6997 (1999).
\bibitem{Guo_preprint} J. Taylor, H. Guo, and J. Wang, cond-mat/0007176.
\bibitem{Joachim_98}C. Joachim, J. K. Gimzewski, and H. Tang,
Phys. Rev. B {\bf 58}, 16407 (1998).
\bibitem{Gaussian} {\em Gaussian 98}, Gaussian Inc., Pittsburgh, PA,
1998.
\bibitem{Ordejon} P. Ordej\'on, E. Artacho, and J. M. Soler, Phys.
Rev. B {\bf 53} R10441 (1996).
\bibitem{Martin} L. Mart\'{\i}n-Moreno and J. A. Verg\'es, Phys. Rev.
B {\bf 42}, 7193 (1990).
\bibitem{note1}The nearest-neighbor hopping elements of the tight-binding
model for the Bethe lattice can be obtained in multiple ways. We have
chosen here to extract them
from a DF calculation of a dimer at the bulk nearest-neighbor distance 
of the considered electrode. The diagonal elements are shifted at
will to reproduce the desired Fermi level.
\bibitem{Becke} A. D. Becke, J. Chem. Phys. {\bf 98}, 5648 (1993).
\bibitem{note2} A structural relaxation is expected to occur, but
this is not relevant to show the reliability of the method.
\bibitem{note3}Work in progress.
\bibitem{Fasel} R. Fasel {\em et al.}, Phys. Rev. Lett. {\bf 76}, 4733
(1996).

\end{references}
\end{document}